# Understanding the Geometric Diversity of Inorganic and Hybrid Frameworks through Structural Coarse-Graining

Thomas C. Nicholas, Andrew L. Goodwin, and Volker L. Deringer*

*Department of Chemistry, Inorganic Chemistry Laboratory, University of Oxford, Oxford OX1 3QR, UK*

**Abstract:** Much of our understanding of complex structures is based on simplification: for example, metal–organic frameworks are often discussed in the context of "nodes" and "linkers", allowing for a qualitative comparison with simpler inorganic structures. Here we show how such an understanding can be obtained in a systematic and quantitative framework, by combining atom-density based similarity (kernel) functions and unsupervised machine learning with the long-standing idea of "coarse-graining" atomic structure. We demonstrate how the latter enables a comparison of vastly different chemical systems, and use it to create a unified, two-dimensional structure map of experimentally known tetrahedral $AB_2$ networks – including clathrate hydrates, zeolitic imidazolate frameworks (ZIFs), and diverse inorganic phases. The structural relationships that emerge can then be linked to microscopic properties of interest, which we exemplify for structural heterogeneity and tetrahedral density.

## Introduction

Establishing links between chemical structure and function is a key requirement for developing new materials. The synthetic exploration of solid-state structural space has been documented in extensive databases,[1] and high-throughput computations and structure prediction are poised to accelerate it further.[2] In an aim to navigate this vast space, lower-dimensional representations have been proposed, such as 2D "maps" with chemically informed coordinates, aiming to identify promising synthesis targets.[3]

With machine learning (ML) approaches currently burgeoning in materials chemistry,[4] it is natural to ask whether they might help in this regard. ML algorithms can handle very large datasets, but are (deliberately) chemically agnostic, and it is not *a priori* clear whether they will discover the same relationships that a trained chemist identifies just by eye. In this context, "unsupervised" ML means that information is sought from a set of data without labels[5] – for example, from a mathematical representation of the atomic structure, for which reliable tools are now available.[6]

One such structural representation is the Smooth Overlap of Atomic Positions (SOAP) similarity function, or *kernel*.[6c] Initially used for fitting machine-learned force fields,[7] it was suggested in 2016 that SOAP can be used for visualising chemical space.[8] Applications to date include known and hypothetical ice structures,[9] the $TiO_2$ polymorphs,[10] molecular crystals,[11] hypothetical zeolites,[12] and Li-ion battery anodes.[13] Once a SOAP-based structure map has been created, it can be used, *e.g.*, to select the most representative motifs for computational spectroscopy.[14] Very recently, zeolites were studied with SOAP-based maps and assessed regarding their topology and synthesisability.[15] The advantages of SOAP over "classical" descriptors, in the context of zeolites, have been emphasised in that work: SOAP not only recovers information normally related to distances, angles, and ring sizes in zeolites, but also subtleties beyond.[15] However, all these applications have been limited to one specific system or to its immediate chemical vicinity.

Here, we generalise this approach across vastly different families of chemical structures, and thereby develop a framework in which geometric diversity can be quantified, visualised, and better understood. The key enabling step is the realisation that a density-based metric such as SOAP can be applied equally well to coarse-grained and uniformly scaled representations of structures as to the structures themselves: this allows us to compare compounds with inherently different chemistries and bond lengths. With a long-term aim of discovering (and, ultimately, exploiting) structural relationships, we focus this proof-of-concept study on one notoriously diverse and important family of inorganic and hybrid frameworks: namely, the $AB_2$-type networks with tetrahedral-like $[AB_4]$ environments.

## Results and Discussion

We start by noting that whilst comparisons across $AB_2$ structures have been eminently useful,[16] they have normally been limited to individual aspects either of the structure (say, the A–B–A angles) or topology (thereby removing subtleties of the structure itself). For example, zeolitic imidazolate frameworks (ZIFs), such as ZIF-8 (Figure 1a),[17] have been discussed in terms of the analogy to Si–O–Si angles in $SiO_2$ polymorphs.[18] We now use a computer algorithm for the same task: placing "dummy" atoms at the midpoint between those (nitrogen) atoms that connect to the $Zn^{2+}$ centres, as shown in Figure 1a. (Note that this is not the same as the centre of mass of the *entire* linker, which would distort the resulting angles for larger ligands, such as benzimidazolate.)

A classical inorganic example is hydro-sodalite (Figure 1b).[19] In this case, we need to remove intra-framework $Na^+$ ions and water from consideration. We also discard the chemical distinction between two different cation sites – now represented by a single "A" dummy atom – but retain any geometric differences in their local environments. This idea of increasing the granularity of the structure is in analogy to how coarse-graining approaches are used for molecular-dynamics simulations that traverse atomistic and larger scales,[20] and how secondary building units (SBUs) are identified in inorganic solids and metal–organic frameworks.[21] We refer to the resulting approach, including removal of guests, coarse-graining, and re-scaling, as "cg-SOAP" in the following.

---

*E-mail: volker.deringer@chem.ox.ac.uk

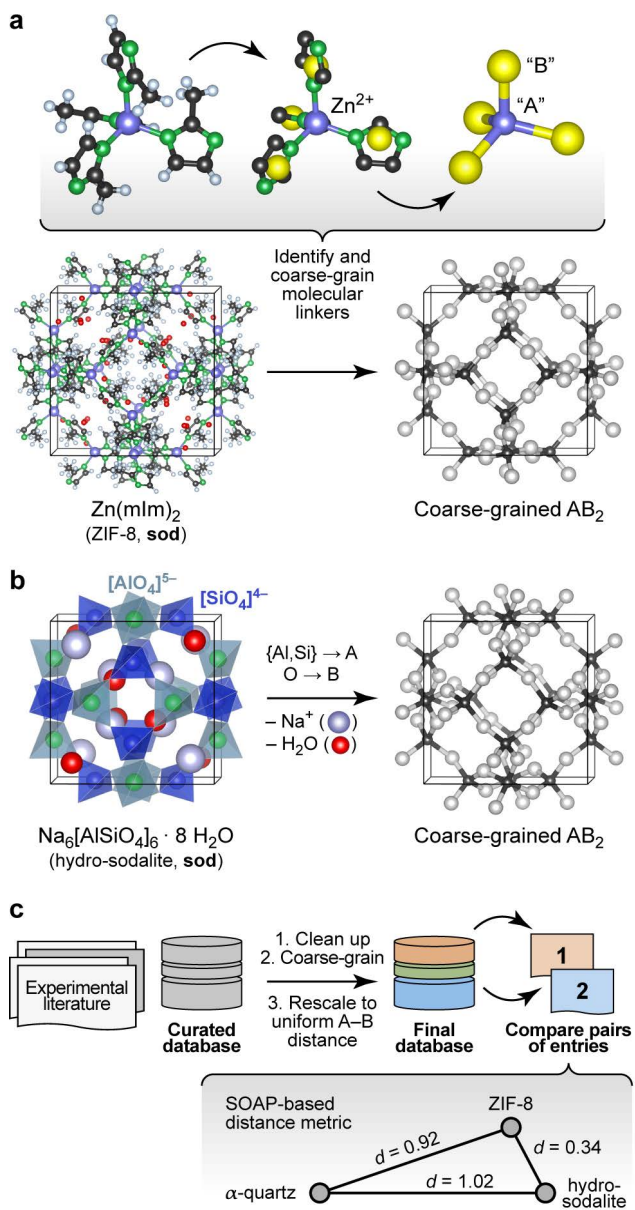

**Figure 1.** Understanding complex tetrahedral inorganic and hybrid structures by reducing them to the underlying AB$_2$ networks ("coarse-graining"). (**a**) The prototypical zeolitic imidazolate framework, ZIF-8,[17] can be reduced by placing dummy atoms (represented by yellow spheres) at the midpoint of the N···N contact inside a single methyl imidazolate (mIm) linker. The resulting simplified ("coarse-grained") structure contains an "A" atom for each Zn$^{2+}$ position, and a "B" atom for each linker: we obtain an open AB$_2$-type structure with four- and six-membered rings (**sod** topology label; Ref. [28]). (**b**) The crystal structure of the inorganic mineral hydro-sodalite[19] is based on the same framework topology. To illustrate this relationship, we remove the (partly occupied) Na sites and the water molecules within the framework, and we reduce the Al and Si cation sites to a single "A" atom. This way, we arrive at a representation that looks very similar to that of ZIF-8 above. There are still differences in the orientation of the individual tetrahedra, and characteristically different absolute A–B distances, which need to be re-scaled for proper comparison. (**c**) Overview of the workflow in the present study, with database building, processing, and then analysis. The inset illustrates the concept of SOAP-based distances, $d$, for a set of three structures: ZIF-8 and hydro-sodalite (shown above) are quite similar in their coarse-grained and rescaled representations; α-quartz is very different from both. Structures were visualised using VESTA.[29]

To test this idea on a much wider basis of experimentally validated structures, we assembled a dataset which includes diverse families of AB$_2$-like materials, including zeolites, ices, and chain-like inorganic structures such as BeCl$_2$. All entries were manually verified. Among the data sources, we point out a review article on ZIFs by Yaghi and co-workers,[18a] a report on cadmium-based frameworks ("CdIFs") by Tian *et al.*,[22] and a study of polymorphism in Zn(CN)$_2$ by Chapman and co-workers.[23] More structures are collected from the Cambridge Structural Database[1b] and the IZA Database of Zeolite Structures.[1e] The full data and references are given as Supporting Data.

Once the coarse graining is done, one key step remains before these very different chemistries can be compared using SOAP: we re-scale the structures to a uniform minimum bond length[24] – an idea that originated in the field of chemical topology.[25] We use the openly available DScribe implementation[26] of SOAP[6c] coupled to ASE.[27] The workflow on which the following analysis is based is shown in Figure 1c.

The SOAP kernel is a similarity measure between two atomic environments, $k(\alpha, \beta)$,[6c] on a scale from 0 (entirely dissimilar) to 1 (identical up to a given cut-off, chosen here to include next-nearest neighbours). Averaging over all A-site environments $\alpha$ in the $i$-th unit cell in our database and $\beta$ in the $j$-th, we obtain a *per-cell* similarity, $\bar{k}(i,j)$. With this, one may then define a distance as

$$d_{ij} = \sqrt{2 - 2\bar{k}(i,j)}$$

to satisfy the triangle inequality (Figure 1c).[8]

We now progress to a much larger structure map that represents distances between many configurations. For this, we use a basic unsupervised ML approach, multi-dimensional scaling (MDS) – a projection into a 2D space, which can be coupled to SOAP.[10,14] Our map is shown in Figure 2 and spans all entries of our manually curated database, classified according to inorganic (*e.g.*, SiO$_2$ polymorphs), molecular (*e.g.*, ice networks), and tetrahedral hybrid networks, *viz.* ZIFs and related cadmium-, boron-, or other cation based tetrahedral imidazolate frameworks ("TIFs"). We follow the naming conventions in the existing literature, accepting that the abbreviations will not always be entirely unambiguous – *e.g.*, for cadmium-based species: Cd(Im)$_2$-**dia-c** was labelled as a "ZIF" in Ref. [18a], whereas Cd(mIm)2-**sod** was initially reported as "CdIF-1" shortly thereafter.[22]

In the 2D space of Figure 2, structures that are similar appear close together, and structures that are dissimilar are further apart. Some material classes are widely distributed throughout the space spanned by the map, with the widest *absolute* distribution found for the zeolites ("+"). Hybrid frameworks (blue symbols) occupy some of this space, but distinctly not all of it; SiO$_2$ polymorphs and disordered ices (such as the common ice-I) are widely spread as well, whereas ordered ices are clustered closely together in the bottom left area. In addition to the absolute distribution across the map, we may quantify the *relative* distribution for each materials class, by which we mean the standard deviation of how far points are from their respective centre of mass, normalised such that the SiO$_2$ polymorphs have a relative distribution of 1.0. ZIFs (zeolites) attain values of 1.20 (1.04), respectively. On the other hand, the ordered ices have a relative distribution of only 0.06, consistent with lower structural flexibility in their strongly directional hydrogen-bonded networks.

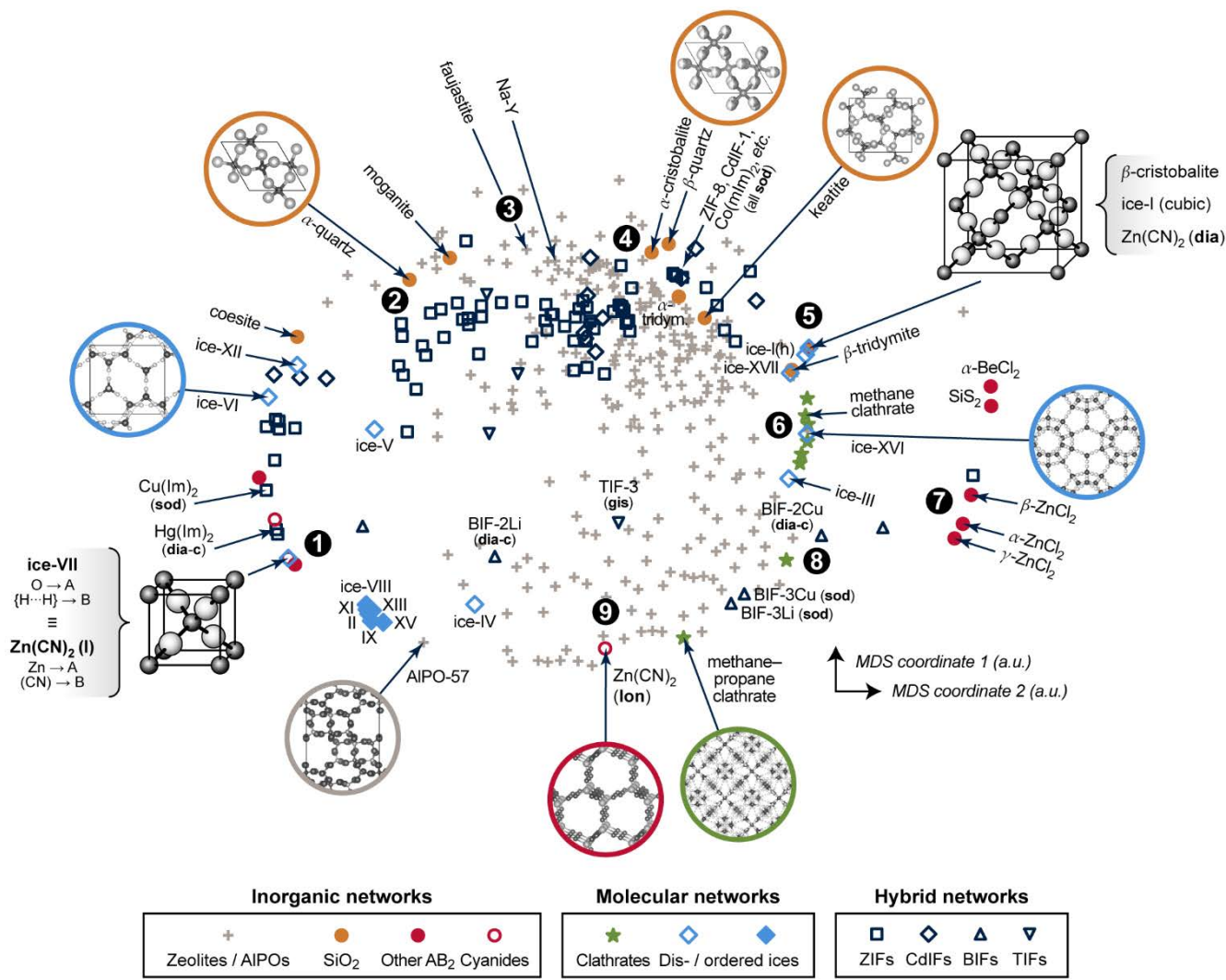

**Figure 2.** A two-dimensional map for inorganic and hybrid tetrahedral structures. The closer two points are, the more similar the corresponding structures, and vice versa. This visualisation is based on a structural dissimilarity (distance) metric, using the SOAP kernel to compare coarse-grained and re-scaled structures (*cf.* Figure 1c), and on embedding by multi-dimensional scaling (MDS). Different symbols are used for the various types of inorganic, molecular, and hybrid networks that are all part of our database. Points of interest are marked as **1**, **2**, and so on, and discussed in the text in this order.

We now walk through this map in clockwise direction, having labelled some more specific locations of interest with boldface numbers. In the lower left part, there is a point where two structures coincide exactly (**1**). One is disordered ice-VII, where we reduce the O–H···H–O bridge (with both hydrogen sites half-occupied) to an A–B–A link. The other is the ambient polymorph of zinc cyanide, for which we also reduce the Zn···C≡N···Zn motif to a symmetric A–B–A link because of head-to-tail orientational disorder of the CN⁻ linkers. Both phases are based on the same anticuprite structure, with no internal degrees of freedom; hence the two points coincide perfectly. LiCo(CO)$_4$ adopts a lower-symmetry variant of the same structure type,[30] with the CO ligand closer to Co than Li – its midpoint being shifted along ($x, x, x$) from $x$ = 0.25 to 0.241. That structure is therefore almost, but not exactly, in the same location on the cg-SOAP map.

Moving up past other disordered ices, the silica polymorphs begin to appear in the upper left part of the map in Figure 2. We illustrate α-quartz, the stable form (**2**). Not many hybrid frameworks (blue) are found in its immediate vicinity, from which we infer that its particular geometry is relatively unusual in the wider context of AB$_2$ networks. We move clockwise past more open framework structures, *viz.* the faujastite and Na-Y cages (**3**), and we find β-quartz near the top of the map (**4**). In the immediate vicinity, there is then a rather largely populated cluster of ZIFs and related structures. Of note is the cadmium-based framework, CdIF-1, which has **sod** topology (*cf.* Figure 1b), and is therefore located alongside other sodalite-type ZIFs and the zeolite itself.

β-cristobalite is another high-symmetry structure with no internal degrees of freedom, located in the upper right part of the map (**5**). After coarse-graining and re-scaling, cubic ice-I and **dia**-Zn(CN)$_2$ occupy exactly the same location; hexagonal ice-I is very close. We find a region of clathrate hydrates (**6**), related to the "empty" frameworks of the low-density ices III and XVI, reflected in close proximity in the cg-SOAP map. Separated clearly from the main area of the map, there is then an "island" of inorganic structures on the right-hand side (**7**): *e.g.*, SiS$_2$, which features chains of edge-sharing tetrahedra, very different from the SiO$_2$ polymorphs in which all tetrahedra are corner-sharing.

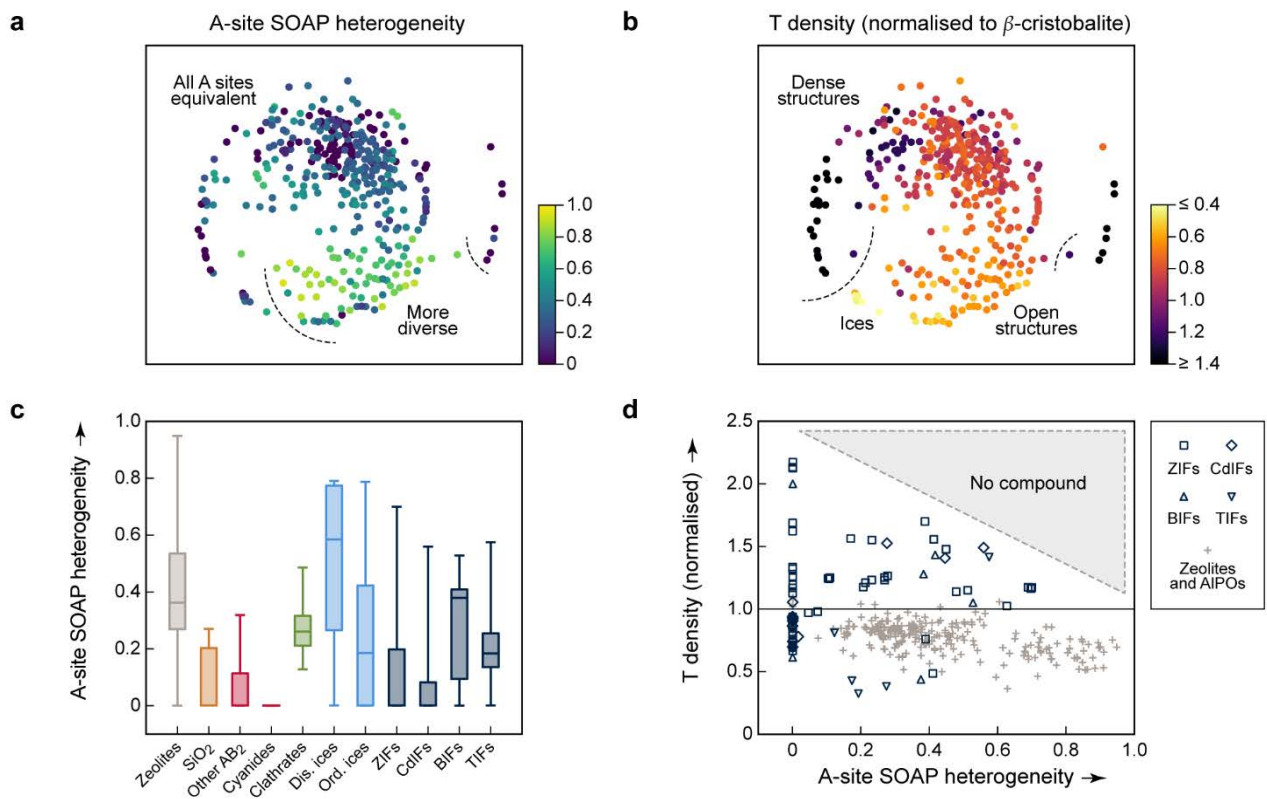

**Figure 3.** Structure–property correlations in diverse tetrahedral networks analysed with our methodology. (**a**) A-site SOAP heterogeneity (that is, a measure for how dissimilar cationic environments are within a given structure), colour-coded on the 2D map from Figure 2. (**b**) Tetrahedral ("T") density, given relative to β-cristobalite, colour-coded on the same map. (**c**) A more quantitative analysis of the A-site SOAP heterogeneity, in which the data have now been collected according to the different categories. The box plots indicate the distribution of data: the boxes range from the 25th to the 75th percentile (with a horizontal line indicating the median), and the whiskers indicate the full range of data points. For boxes without a visible horizontal line, the median is zero. (**d**) Connecting both quantities for framework materials and zeolites: the T density for each corresponding entry of our database has been plotted as a function of A-site heterogeneity. There is a class of low-density zeolites ("+") that correlate with large A-site heterogeneity (> 0.6), but dense structures require local homogeneity.

In the lower right part of Figure 2, we observe again more open frameworks. Of note are the boron-based BIFs (**8**), which contain $Li^+$ or $Cu^+$ in combination with $B^{3+}$, and therefore are aliovalent equivalents to ZIFs ($M^{2+}$).[31] We re-iterate that although we reduce the cation sites to a single type of "A" dummy atom, we do retain the *relative* differences in bond lengths around $M^+$ *vs.* $B^{3+}$; therefore, the BIF-3 frameworks are not near other **sod** structures. Finally, near the bottom of the cg-SOAP map in Figure 2, we point out another form of zinc cyanide (**9**), emphasising the large variety of polymorphs that is accessible to a single system.[23] This particular one adopts the same topology as hexagonal ice-I (**Ion**) – but in the $Zn(CN)_2$ structure, the metal⋯cyanide distances are very dissimilar, about 1.6 and 2.0 Å respectively, and the data point is therefore away from ice-Ih in the 2D map of Figure 2. In the context of cyanides, we mention the even larger structural diversity in Prussian blue analogues:[32] this exemplifies a limit of our method in that it needs discrete positions for the "B" grains, and it cannot capture longer-range correlated disorder.[33]

An important property of a materials map is that it should be able to be correlated with relevant properties.[3] The first quantity for which we test this question is again concerned with structural diversity. In Figures 1c and 2, we had used an averaged metric to compare different unit cells with one another – but SOAP can also be used to compare individual atoms within one and the same structure. We may therefore use it to assess the question of how diverse the different A-sites in any given structure are, which we call "A-site SOAP heterogeneity": a value of zero means that all A-site environments (normally, metals) are geometrically equivalent, and a higher value indicates a higher degree of diversity – *e.g.*, in the BIFs, where different aliovalent cationic species occupy the A site, as mentioned above. This information can be visualised in a colour-coded version of our map, which is shown in Figure 3a.

SOAP maps are beginning to be used to identify properties of application interest.[15] In the context of the present work, a central such property is the tetrahedral ("T") density: this is the simplest proxy for possible usefulness in catalysis, because low T densities indicate the presence of voids, which could be used for the absorption, diffusion, and transformation of guest molecules. We show a colour-coded version of our map, illustrating this property, in Figure 3b. Again, there are clearly different regions, evidencing the physical significance of the initially "agnostic" unsupervised ML approach. The two colour-coded maps also show an inherent characteristic of the 2D embedding: it needs to balance all structural aspects, and therefore the very dense networks at the bottom left are close to the very open, ordered ices in Figure 3b. We presume that this is linked to the A-site heterogeneity, which is low in both groups, and prohibits the ices from being in the lower right region with its more diverse A sites (Figure 3a).

The embedding of high-dimensional distances in 2D invariably leads to the loss of some information. It is therefore useful, *in addition* to the map, to now look quantitatively at similarities and properties independent from where a given material is located in the 2D map. We quantify the distribution of A-site SOAP heterogeneities, separately for the different materials classes, in Figure 3c. Some of the $SiO_2$ polymorphs include locally heterogeneous environments (the monoclinic structures of moganite, 0.27, and coesite, 0.21, are of note) – but most of them do not, and neither do most other inorganic $AB_2$ structures. In clathrates, on the other hand, we do not find any fully homogeneous structure (even the minimum value being > 0). Ices are distinct as to whether they are ordered or not. Among the framework materials, CdIFs are the least locally heterogeneous, which is perhaps surprising given the large ionic radius and polarisability of $Cd^{2+}$; BIFs are spread more widely in Figure 3c with the median at almost 0.4, as expected due to the presence of two different cationic species.

Finally, the information content of Figures 3a–b can be combined to study correlations between different property indicators. We do this for the subset of hybrid frameworks and zeolites (Figure 3d). There is a number of fully locally homogeneous structures, mainly composed of the different hybrid framework materials (at *x* = 0), but there are also two distinct regions of heterogeneity (up to *x* = 0.6 and beyond it, respectively), dominated by zeolite structures ("+"). Generally, Figure 3d reveals that all heterogeneous tetrahedral networks studied have low density, and conversely all dense networks are homogeneous; there is a distinct region where no compounds have been experimentally observed, indicated by shading. When aiming to design new low-density materials, one might therefore attempt to introduce and tune A-site heterogeneity. The latter can be achieved experimentally, *e.g.*, by exploiting solid–solution chemistry, both regarding isovalent or aliovalent cations, and combinations of different linkers.

## Conclusions

We have shown how structural relationships across diverse material families can be understood by combining the idea of coarse-graining and scaling atomistic structure with a suitable atom-density based similarity metric (here, SOAP). Our study built on experimentally characterised structures and a carefully curated database of those, but similar approaches may now be extended to even larger sets of data: to hypothetical zeolites,[34] hybrid perovskites,[35] or to MOFs with non-tetrahedral structures,[36] for example. In regard to visualisation tools, we used one of the simplest (*viz.*, MDS), which already leads to appreciable results, but one might seamlessly couple our approach to other, more involved visualisers such as the popular sketch-map scheme[37] or *t*-stochastic neighbour embedding[38] which are also beginning to be used with SOAP.[8,15] To this end, our database of all coarse-grained representations will be made openly available online upon publication of this work, with the hope to enable future work in the community.

## Acknowledgements

We are grateful for support from the European Research Council (ERC Advanced Grant 788144, to A.L.G.) and the Leverhulme Trust (Early Career Fellowship, to V.L.D.).